\documentclass[submission, copyright]{eptcs}
\providecommand{\event}{F-IDE 2019} % Name of the event you are submitting to
\usepackage{underscore}           % Only needed if you use pdflatex.

\usepackage[utf8]{inputenc}
\usepackage[T1]{fontenc}
\usepackage{graphicx}
\usepackage{wrapfig}

\newcommand*\circled[1]{
	\hspace{-1mm}\raisebox{.5pt}{\textcircled{\raisebox{-1pt} {\hspace{0.05mm}\small#1}}}}

\title{Automated Deductive Verification for Ladder Programming}
\author{Denis Cousineau
\qquad \qquad
David Mentré
\institute{Mitsubishi Electric R\&D Centre Europe (MERCE)\\ Rennes, France}
\email{\{d.cousineau,d.mentre\}@fr.merce.mee.com}
\and
Hiroaki Inoue
\institute{Mitsubishi Electric Corporation\\ Amagasaki, Japan}
\email{Inoue.Hiroaki@ah.MitsubishiElectric.co.jp}
}

\def\titlerunning{Automated Deductive Verification for Ladder Programming}
\def\authorrunning{D. Cousineau, D. Mentré \& H. Inoue}
\begin{document}
\maketitle

\begin{abstract}
Ladder Logic is a programming language standardized in IEC 61131-3 and widely used for programming
industrial Programmable Logic Controllers (PLC).  A PLC program consists of inputs (whose values are
given at runtime by factory sensors), outputs (whose values are given at runtime to factory actuators),
and the logical expressions computing output values from input values. Due to the graphical form of
Ladder programs, and the amount of inputs and outputs in typical industrial programs, debugging such
programs is time-consuming and error-prone. We present, in this paper, a Why3-based tool research prototype we
have implemented for automating the use of deductive verification in order to provide an easy-to-use
and robust debugging tool for Ladder programmers.
\end{abstract}

%%%%%%%%%%%%%%%%%%%%%%%%%%%%%%%%%%%%%%%%%%%%%%%%%%%%%%%%%%%%%%%%%
%%% SECTION

\section{Introduction}

Programmable logic controllers (PLC) are industrial digital computers used as automation controllers of manufacturing processes, such as assembly lines or robotic devices. PLCs can simulate the hard-wired relays, timers and sequencers they have replaced, via software that expresses the computation of outputs from the values of inputs and internal memory. Ladder language, also known as Ladder Logic, is a programming language used to develop PLC software. This language uses circuits diagrams of relay logic hardware to represent a PLC program by a graphical diagram. This language was the first available to program PLCs. It is now standardized in IEC 61131-3 \cite{iec:61131} standard among other languages but is still widely used and very popular among technicians and electrical engineers. 

In conventional development of software, a great part of the development time is dedicated to debugging. Debugging programs is crucial in the case of Factory Automation (FA) since bugs in factories can be extremely expensive in terms of human and material damages, and plant downtime. Debugging a Ladder program is particularly difficult, time consuming and costly. Bugs can be depicted as the violation, at some point of a program, of some property concerning values of inputs/outputs and local memory of the program. The objective of debugging consists in detecting those property violations before running the code in production, i.e. finding initial values of inputs and internal memory that lead to a property violation, when executing the program. Since it is almost impossible (and way too costly) to check all possible executions of a program, the usual method consists in developing and running some tests (i.e. executing the program on a particular initial configuration and check its behavior). In industry, tests used to be run directly in the factory, which is very costly and risky. Nowadays, most of the tests are run on a software simulation, but some are often still run in the factory for a last check of the program behavior in real conditions of use, or for bypassing the difficulty to simulate particular sequences of inputs. Even when run on a software simulation, tests-based processes are still time-consuming and cannot be exhaustive.

On the other hand, some research work has been done concerning formal analysis of Ladder programs. Most of this work \cite{884356} \cite{7295624} \cite{7925390} \cite{overview-MC}  concerns the verification of temporal properties of Ladder programs (a Ladder program being continuously executed in the PLC), and uses different model checking techniques. Some other work used deductive verification to detect data races \cite{Su:CSD-97-969} and prove safety properties \cite{roussel:hal-00356881} (with some temporal aspects) of Ladder programs. Model-checking techniques are limited by the state explosion problem they face when addressing real-world problems. On the contrary, deductive verification may give full confidence in the obtained results but may also prevent from a full automatization of the process (in terms of proof automation and specification formalization).%+ ref Spark

Our objective in this work was to make a proof of concept of an easy-to-use and robust tool for debugging Ladder programs, both increasing the quality of the code and decreasing the time to deploy, a crucial point in the context of Industry 4.0, in which assembly lines are more often reconfigured, hence code evolves frequently. We had to determine a good tradeoff to offer a high level of automation, together with providing a strong confidence in the given results (in particular when no bug is found). The solution we chose is similar to what SPARK/GnatPro \cite{hauzar:hal-01314885} has done for Ada code. For the easy-to-use part, we targeted a full automated tool, so that it could be used by regular engineers with no needed knowledge in formal methods. We also focused on the information given to the programmer when an error is found, for easing errors fixing. In order to obtain a fully automated and complete tool, we targeted to detect runtime errors by the mean of deductive verification. We focused on runtime errors like integers overflows, divisions by zero, violations of Ladder instructions' pre-conditions, etc... 

 We based our prototype implementation on the Why3 platform \cite{bobot:hal-00967132}. Why3 offers an expressive formalization language, an efficient Weakest-Precondition (WP) calculus \cite{Dijkstra:1997:DP:550359} implementation and a rich API to send the obtained verification conditions to several automated solvers. Moreover, with its \textit{labels} mechanism, Why3 allowed us to keep code information during the whole automatic process, for providing rich and useful information to the programmer in case a bug is detected.

%%%%%%%%%%%%%%%%%%%%%%%%%%%%%%%%%%%%%%%%%%%%%%%%%%%%%%%%%%%%%%%%%
%%% SECTION

\vspace{-3mm}

\section{Ladder Logic}
Ladder Logic is a graphical programming language using relay logics diagrams to represent a PLC program. A Ladder program takes inputs values (\textit{contacts}) that correspond to the fact that physical relays are wired, not wired, pulsing (rising edge) or downing (falling edge) and other values stored in the internal memory of the PLC (booleans, integers, floating point, strings, etc...). A Ladder program can output boolean values to the physical relays of the factory (\textit{coils}) or it can call instructions, that may modify the values of the internal memory of the PLC (\textit{devices}). Graphically, contacts are located at the left of the diagram. They can be combined in a serial way or in a parallel way (the obtained value is then the conjunction, resp. the disjunction of the two contacts values). Coils and instructions are activated when the combination of contacts at their left gives a \textit{wired} value,  and they can also be parallelized (in that case, there are either all activated or all deactivated). A line with contacts, coils and instructions is called a \textit{rung}, and a program (a \textit{diagram}) is composed of several rungs. 

\begin{figure}[!h]
\includegraphics[width=16cm]{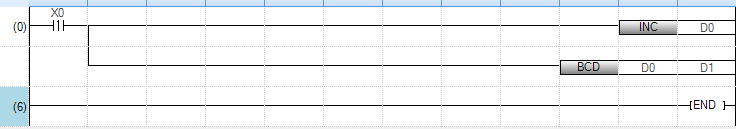}
\vspace{-7mm}
\caption{Ladder program example}
\label{exladder}
\end{figure}

\noindent Figure \ref{exladder} depicts a very simple Ladder example. This program has one contact \texttt{X0}, and when the physical relay corresponding to that input is activated, the program calls instruction \texttt{INC} that increments the value of its device argument \texttt{D0}, and then calls instruction \texttt{BCD} with \texttt{D0} and \texttt{D1} devices as respectively input and output arguments. The \texttt{BCD} instruction converts a 16 bits integer into a 16 bits BCD (Binary-Coded Decimal) integer. The 16 bits BCD format represents 4 digits decimal numbers, using 4 bits to represent each of the 4 digits. It is typically used for display purpose. Since this format can only represent 4 digits decimal numbers, the \textit{BCD} instruction raises an error when it is called on a device value that does not belong to interval $[0;9999]$. This is typically the kind of runtime errors we want to detect with the tool we developed. Regarding this example, we were also interested in overflows that could occur when calling instruction \texttt{INC}.
The example we present is very simple but typical industrial programs we had access to have hundreds of lines, hundreds of inputs, devices and outputs, and dozens of instructions calls. As a last point, such a Ladder program is executed cyclically in a synchronous way: first inputs are read, then the program is executed and eventually outputs are written. One single execution of the program is called a \textit{scan}.

%%%%%%%%%%%%%%%%%%%%%%%%%%%%%%%%%%%%%%%%%%%%%%%%%%%%%%%%%%%%%%%%%
%%% SECTION

\vspace{-3mm}

\section{Modelling Ladder in Why3}

\begin{wrapfigure}{r}{0.7\textwidth}
  \vspace{-8mm}
\begin{center}
    \includegraphics[width=0.68\textwidth]{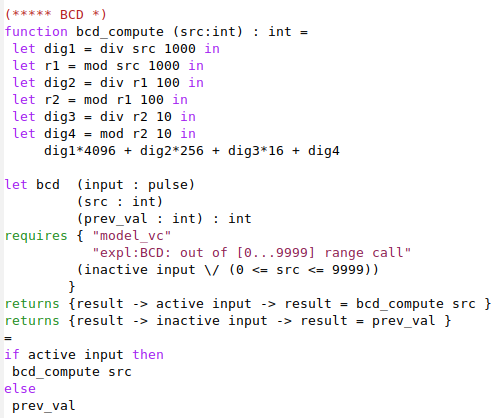}
  \end{center}
\vspace{-7mm}
  \caption{Why3 formalization of Ladder BCD instruction}
  \label{why3lib}
\end{wrapfigure}

We chose not to model the temporal/cyclic aspect of execution of Ladder programs, but only one scan in order to detect error scenarios, i.e. values of inputs and devices at scan beginning (before execution) that may lead to a runtime error. We developed a library of Ladder instructions formalizations.
We depict here the formalization of the \texttt{BCD} instruction. This formalization is composed of two functions. The first one, \texttt{bcd_compute}, computes the 4 digits of an decimal integer argument and returns the decimal value of the BCD representation of those 4 digits. The second one, \texttt{bcd} takes three arguments: \texttt{input} is the wiring value of the line to which the instruction is connected, \texttt{src} is the value of the input device, and \texttt{prev_val} is the value of the output device before execution of the instruction. The \texttt{requires} pre-condition states that either \texttt{input} does not activate the instruction or \texttt{src} must belong to interval $[0;9999]$. You can notice the two strings \textit{labels} in the pre-condition. The first allows asking solvers to find a counter-example if they cannot prove the verification conditions associated with that pre-condition. The second one allows keeping semantic information during the whole process, in order to give back this information to the programmer in case an error scenario is found. The \texttt{returns} post-condition states that the function returns the previous value of the output device when the instruction is not activated, and the actual BCD computation otherwise.

We developed around fifty such formalizations of Ladder instructions in order to run our tool prototype on the industrial program samples we had access to. Then our translation of Ladder programs to Why3 models consists in translating on-the-fly the logical expressions that correspond to coils and their combinations, and combine them with calls to the instructions formalizations of our library, using a single-state-assignment \cite{Rosen:1988:GVN:73560.73562} transformation to handle the iterative aspect of Ladder programs.

%%%%%%%%%%%%%%%%%%%%%%%%%%%%%%%%%%%%%%%%%%%%%%%%%%%%%%%%%%%%%%%%%
%%% SECTION

\section{Prototype architecture}

\begin{wrapfigure}{r}{0.51\textwidth}
  \vspace{-13mm}
\begin{center}
    \includegraphics[width=0.5\textwidth]{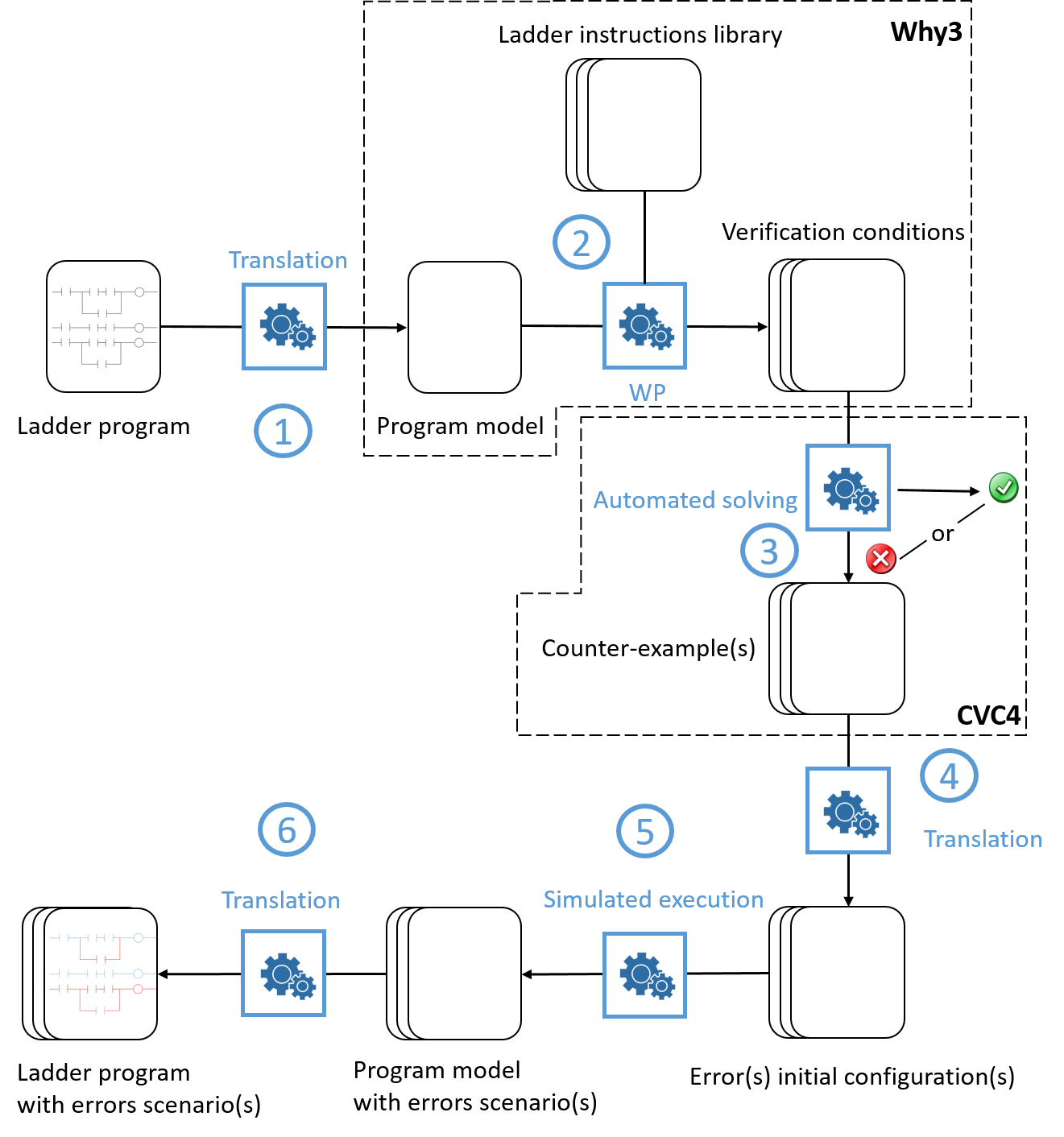}
  \end{center}
\vspace{-6mm}
  \caption{Prototype architecture}
\end{wrapfigure}

Our tool automatically  \circled{1} translates Ladder programs into Why3 modules that refer to the instructions formalizations described in the previous section. We implemented our own library to produce Why3 text files, to help the reuse of generated modules. During the translation, we use \textit{labels} to keep information on code location of instruction calls to give improved feedback to the programmer when an error is found. 
\newline
Then we use Why3's WP calculus to \circled{2} compute verification conditions that correspond to pre-conditions of instructions, and use Why3 API to \circled{3} send those verification conditions to SMT-solver CVC4 \cite{Barrett:2011:CVC:2032305.2032319} (we chose CVC4 for its overall good performances and its ability to generate counter-examples when a verification condition cannot be proved). Couter-examples are then \circled{4}  interpreted as initial values of the original program and simulated execution \circled{5} recomputes, from those inital values, all the intermediate values of devices, wires, etc... from the beginning of the program to the location where the error occurs.  Finally, we \circled{6} provide a graphical feedback to the programmer, with those intermediate values information and informations concerning the error the tool found.

%%%%%%%%%%%%%%%%%%%%%%%%%%%%%%%%%%%%%%%%%%%%%%%%%%%%%%%%%%%%%%%%%
%%% SECTION

\section{Graphical user feedback}

We implemented a proof of concept of a graphical interface which gives back to the programmer information about found bugs, in an easy-to-understand manner. The aim of such an interface is to be directly integrated in Ladder IDEs. 
We based our prototype implementation on the Ocsigen web framework \cite{balat:hal-00150444} which allowed us to quickly prototype a web-based graphical interface displaying information coming from our tool prototype implemented in OCaml. We identified three pieces of information that should be displayed to the programmer when a bug is found: the \textit{error location} (where the error occurs), the \textit{error reason} (why the error occurs) and the \textit{error scenario} (when the error occurs). 
The error location is encoded during the on-the-fly translation from the Ladder program to the Why3 model: to each instruction call is attached a label which contains its location in the original source code. This label is propagated during the WP calculus, appears in the verification condition sent to the automated solver and comes back in the counter-example the solver gives when it finds one.    
The error reason is encoded in the Why3 instructions library as shown in figure \ref{why3lib}. It is attached, with a \texttt{"expl:"} label, to pre-conditions of Ladder instructions, and is propagated during the whole process, like code locations labels. 
As explained in the previous section, the error scenario consists in the initial and intermediate values that lead to the error. It is re-computed from solvers' counter-examples, and is expressed with colors for wiring values (blue when a wire is active, grey otherwise) and figures for other values above the corresponding devices.

\begin{wrapfigure}{r}{0.66\textwidth}
  \vspace{-7mm}
\begin{center}
    \includegraphics[width=0.65\textwidth]{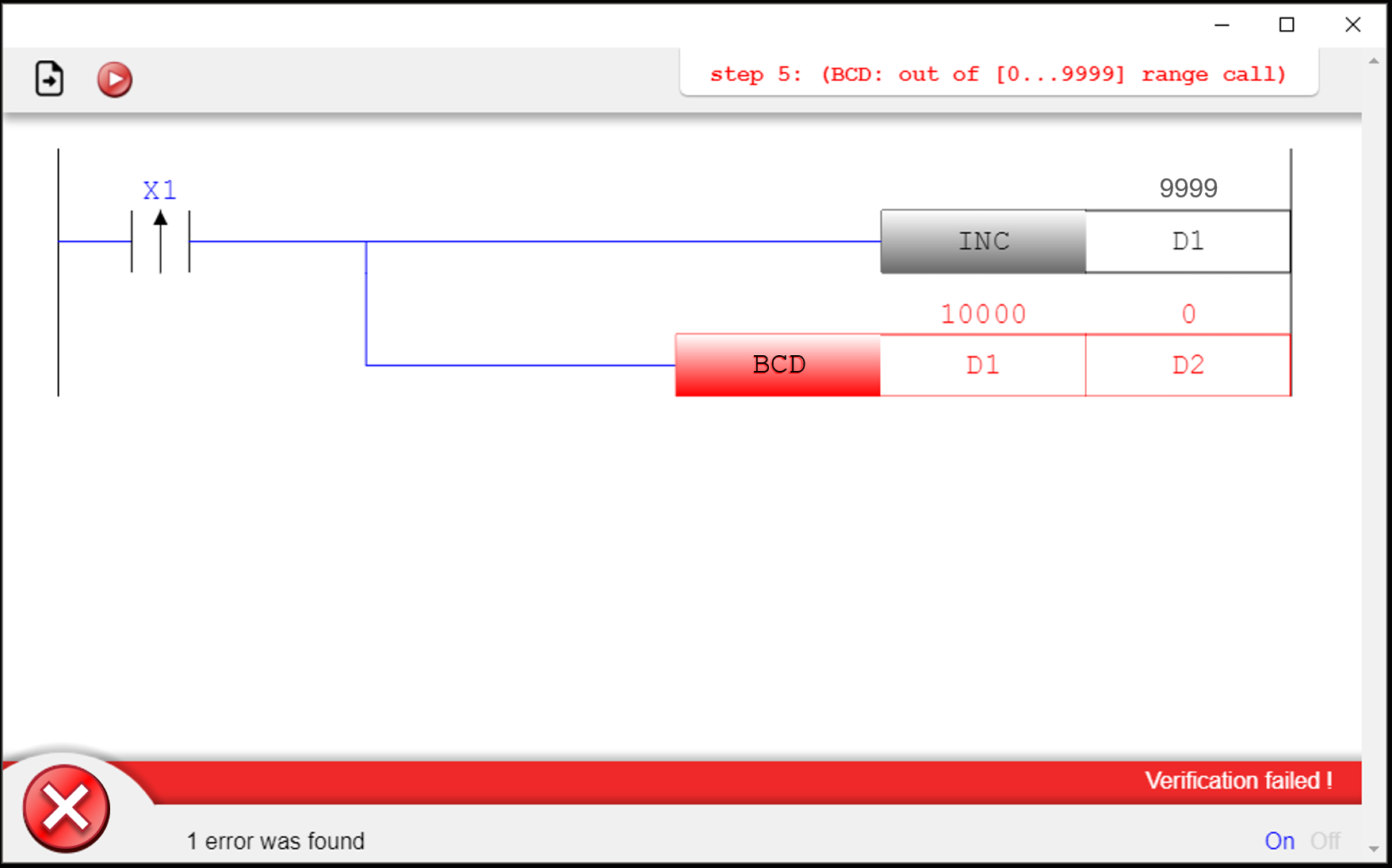}
  \end{center}
\vspace{-6mm}
  \caption{Graphical feedback}
\label{gui}
\end{wrapfigure}

\noindent Figure \ref{gui} shows a screenshot of this graphical interface. This is what our prototype returns when run on example of figure \ref{exladder}. In this case, the interface states the errors occurs at BCD instruction call location (it is colored in red). It also states that the error reason is an out-of-range call. And it gives the error scenario: contact \texttt{X1} is active hence colored in blue, then the wires at its right are also active and colored in blue. After execution of instruction \texttt{INC}, value stored in device \texttt{D1} is $10,000$, which leads to the error, when given as argument of instruction \texttt{BCD}.

\noindent We believe that this interface, somehow inspired by what already exists in Ladder simulation software, in particular for the colors, may be very useful for debugging Ladder programs, in particular when they reach a critical size with hundreds of rungs, inputs, devices, outputs, etc... Indeed, it is much easier to understand why an error occurs with this kind of interface, than when using tests, in which case only the initial configuration of the program is given. 

%%%%%%%%%%%%%%%%%%%%%%%%%%%%%%%%%%%%%%%%%%%%%%%%%%%%%%%%%%%%%%%%%
%%% SECTION
\vspace{-3.5mm}

\section{Performances}
Our proprietary prototype is implemented in OCaml, in about $13,000$ lines of code, including $3,000$ lines for our library to produce Why3 textual files and $4,000$ lines for the graphical user interface. We made some optimization effort concerning the modelization of Ladder language in Why3, in order to obtain a fully automated and fast process. But we made no optimization in our OCaml code, and even did not parallelize the calls to SMT-solvers. Nonetheless our protype has already pretty good performances. We ran our prototype tool on an industrial code sample with $1,657$ \textit{steps} (i.e. contacts, coils and instruction calls), among which three instructions calls could lead to an error. On a virtualized Ubuntu 18.04, running in VirtualBox 5.2, under Windows 10 on a Intel Core i7-7500U 2.70 GHz laptop, it takes only 3 to 4 seconds for our prototype to answer. Almost all the time is taken by CVC4 (about one second for each of the three verification conditions that are handled sequentially). This gives us confidence in the fact that the technology and architecture we chose are relevant for the implementation of a real industrial tool.

%%%%%%%%%%%%%%%%%%%%%%%%%%%%%%%%%%%%%%%%%%%%%%%%%%%%%%%%%%%%%%%%%
%%% SECTION

\vspace{-2mm}
\section{Conclusion}
The objective of this work was to make a proof of concept of a formal methods-based debugging tool for industrial Ladder programs. For such a debugging tool to be incorporated in an industrial process, we think that it should be transparent and bring strong added value to the user.
 First, our proof of concept shows that such a formal debugging tool can be transparent:
it needs no specific knowledge since all the process is fully automatic (Ladder programmers do not need to write a formal specification, and even less a model of their codes);
it is very fast so it may be run during the programming phase of the development process and not in a separated phase;
it may be fully integrated in a Ladder IDE as our GUI prototype shows. 
Second, our proof of concept shows the added value such a tool could have in regard to current debugging tools: 
 our prototype can give back to the programmer very precise and useful information when it detects an error (using a intelligible interface); 
 and the deductive verification technique we used, thanks to the Why3 platform, gives a strong confidence when the tool detects no runtime error (since it is equivalent to test all possible inputs and devices values configurations). 

\noindent A drawback of our prototype concerns the fact that it may raise false positive alarms, since it only considers one scan of the Ladder program. For example in Figure \ref{exladder}, value of device \texttt{D0} may be changed \textit{after} the \texttt{BCD} instruction call, such that value $10,000$ is never reached. Nevertheless, our prototype would still raise an alarm. In future work, we plan to decrease the number of false alarms by considering a few consecutive scans in our Why3 Ladder formalization. 

\noindent Another way to improve our prototype could be to provide some \textit{quickfix}-like mechanisms to programmers. In example of Figure \ref{exladder}, our prototype could propose to the programmer to add automatically, before the \texttt{BCD} instruction call, a line that resets \texttt{D0} when it does not belong to range $[0;9999]$.

%%%%%%%%%%%%%%%%%%%%%%%%%%%%%%%%%%%%%%%%%%%%%%%%%%%%%%%%%%%%%%%%%
%%% SECTION

\vspace{-5mm}
\bibliographystyle{eptcs}
\bibliography{biblio-cut}

%%%%%%%%%%%%%%%%%%%%%%%%%%%%%%%%%%%%%%%%%%%%%%%%%%%%%%%%%%%%%%%%%
%%%%%%%%%%%%%%%%%%%%%%%%%%%%%%%%%%%%%%%%%%%%%%%%%%%%%%%%%%%%%%%%%
%%%%%%%%%%%%%%%%%%%%%%%%%%%%%%%%%%%%%%%%%%%%%%%%%%%%%%%%%%%%%%%%%
%%%%%%%%%%%%%%%%%%%%%%%%%%%%%%%%%%%%%%%%%%%%%%%%%%%%%%%%%%%%%%%%%
%%%%%%%%%%%%%%%%%%%%%%%%%%%%%%%%%%%%%%%%%%%%%%%%%%%%%%%%%%%%%%%%%
%%%%%%%%%%%%%%%%%%%%%%%%%%%%%%%%%%%%%%%%%%%%%%%%%%%%%%%%%%%%%%%%%

%\nocite{*}
%\bibliographystyle{eptcs}
%\bibliography{generic}
\end{document}